# A New Approach to Determine the Coefficient of Skewness and An Alternative Form of Boxplot

**Ummay Salma Shorna and Md. Forhad Hossain[1]**

**Department of Statistics, Jahangirnagar University, Savar, Dhaka, Bangladesh**

*Abstract*

To solve the problems in measuring coefficient of skewness related to extreme value, irregular distance from the middle point and distance between two consecutive numbers, "Rank skewness" a new measure of the coefficient of skewness has been proposed in this paper. Comparing with other measures of the coefficient of skewness, proposed measure of the coefficient of skewness performs better specially for skewed distribution.

An alternative of five point summary boxplot, a four point summary graph has also been proposed which is simpler than the traditional boxplot. It is based on all observation and give better result than the five point summary.

*Keywords:* Skewness, Rank Skewness, Boxplot, Four Point Summary Graph.

**1.1 Introduction**

In any statistical analysis of a data set, information about the central value, scatterdness and the shape characteristic of the dataset is very important for interpretation, conclusion, prediction and for further advanced analysis purpose. Dealing with symmetric data set and asymmetric data set will not be of similar pattern. Dealing with symmetric data set is much more easier than asymmetric data set, as there are a huge number of statistical tools available in the existing literature for dealing with a symmetric or normally distributed data set. But for asymmetric data set, available statistical tools are very rare.

Hossain and Adnan (2007) have shown that although two datasets may have the same mean and variance, still they may have different shape characteristics such as one may have skewed shape and another may have negatively skewed shape, and even some other may have symmetric shape characteristic. So, checking the asymmetry and determining the amount of skewness by some suitable and appropriate technique is a very important task.

Many renowned statistician have worked to trace the skewnesss and to determine the amount of skewness of a data set and have given the form for determining the coefficient of skewness, e.g.

Fisher had proposed a form which is known as moment based coefficient of skewness such as,

$$SK_F = E\left[\left(\frac{x-\mu}{\sigma}\right)^3\right] = \frac{\mu_3}{\sigma^3}$$

---

[1] Communicating author: Md. Forhad Hossain, email: forhad.ju88@yahoo.com

The main problem of this form is that it is based on moments, and the moments are determined on the basis of mean which is not a proper measures of central value for asymmetric data set. So, this form will not provide us an appropriate measures of the coefficient of skewness.

Another very well known form of measuring the coefficient of skewness suggested by Pearson (1894).This form is based on mean, mode and standard deviation which is as follows,

$$SK_{P_1} = \frac{(mean - mode)}{\sigma}$$

Pearson (1895) had suggested another form for determining the coefficient of skewness which is based on mean, median and standard deviation, such as,

$$SK_{P_2} = \frac{3(mean - median)}{\sigma}$$

Pearson's both coefficients of skewness $SK_{P_1}$ and $SK_{P_2}$ depend on mean as well as standard deviation and both of these mean and standard deviation are not appropriate measure of central value and scatterdness for an asymmetric data set and are also seriously affected by extreme values. So, these form will not provide us an appropriate measures of coefficients of skewness specially when the data set itself is asymmetric.

Sir Arthur Lyon Bowley (1901) had established a measure of skewness which is also called Yule's coefficient of skewness (1912), which is as follows:

$$SK_B = \frac{Q_3 + Q_1 - 2Q_2}{Q_3 - Q_1}$$

Where, $Q_1$, $Q_2$ and $Q_3$, are respectively the first, second and third quartile value of a data set.

Calculation of quartile is complex. Also, it is based on only the middle most fifty percent observation of a dataset. So, ignoring the tailed fifty percent observations, this form do not provide an accurate measure of the coefficient of skewness.

Groeneveld, Meeden and Gin (1984) had described a more generalized form of skewness

$$\gamma(u) = \frac{F^{-1}(u) + F^{-1}(1 - u) - 2F^{-1}(1/2)}{F^{-1}(u) - F^{-1}(u)}$$

The function $\gamma(u)$ satisfies $-1 \leq \gamma(u) \leq 1$ and is well defined without requiring the existence of any moments of the distribution. But it is also based on mean which is influenced by extreme values. So this will not provide an appropriate amount of coefficient of skewness for an asymmetric data set.

They have suggested, another alternative form which is as follows,

$$\frac{(\mu - \nu)}{E(|X - \nu|)}$$

This form also incorporates the mean.

Hossain and Adnan (2007) have suggested a different form for determining the amount of the coefficient of skewness which is as follows,

$$SK_{FA} = \frac{\sum_{i=1}^{n}(x_i - m)}{\sum_{i=1}^{n}|x_i - m|}$$

Where m is the sample median.

Although median is the exact middle value of a dataset and is not affected by any extreme value but it completely ignores the terminal values. In same situation, terminal value(s) may have some important role to deal with the data set and may have a notable impact on measuring the central value of a dataset. So, completely ignoring the information of extreme value in determining the central value, specially when this extreme value do not occur due to some error of the survey or measurement error, is not a wise decision. And hence based on such a central value for measuring the amount of skewness may not provide an appropriate measure of the coefficient of skewness.

In this paper, we have proposed a new approach to determine the coefficient of skewness which is based on midrange.

**1.2 Problem and Motivation**

Skewness essentially measures the relative size of the two tails of a dataset. In some cases, most of the data may condense at the left tail or at the right tail or very close to the central value, but one or more extreme value(s) may wrongly change the size of the tail. There are so many formulae for determining the amount of skewness as we have specified earlier. But all those formulae except Hossain and Adnan (2007) are seriously affected by one or more extreme value(s). Data set may condense to at one tail but due to a single or a few extreme value(s), we may get a misleading result of the measure of the coefficient of skewness.

Symmetry may be explained as equal number of observation at equal distance from the middle or center point of a dataset. Sometimes, there may be some observations with larger distance compared to average distance from the middle point of a dataset. This unusual distance may seriously affect the measure of the coefficient of skewness by using the existing formulae for measuring the coefficient of skewness. All the existing formulae are seriously influenced by the distance of the observation from the center value of the dataset.

Mean is not always an appropriate measures of central value, particularly for an asymmetrically distributed dataset. Although median is the middle most observation of a dataset, still it is not a perfect measure of central value, specially in presence of extreme observation, and also when the distance between two consecutive numbers is irregular. These extreme value(s) may not occur due to measurement error or due to survey error.

As Bowley's measure of the coefficient of skewness is mainly based on boxplot with first, second and third quartile which indicates the degree of skewness through the spacing of different parts. Ignoring the tails 50% observation, it mainly takes into account the middle most 50% observation only. Interpretation of boxplot is more intricate, especially when there is one or more extreme value.

For an asymmetric dataset, arithmetic mean is not a good measure of center value because it is seriously affected by extreme value(s). In this case, though median performs better as a measure of center value but it is based on only the rank of the observations not on the real score of a dataset and is not based on all observations of a dataset. Also, median does not take into account the irregular distance between two consecutive number. In this paper we have proposed an alternative approach for determining the coefficient of skewness.

**2.1 Rank Skewness**

Rank skewness is a new idea of measuring the coefficient of skewness and in this paper, we have suggested an alternative form to measure the amount of the coefficient of skewness, based on this new idea we have termed it as "Rank Skewness". This new approach may play an important role to overcome the problem that we have explained earlier. "Rank Skewness" is thought of as a new era in the measurement of the degree of scatterdness of a dataset. The form of our newly suggested measures of the coefficient of Rank Skewness is as follows,

$$SK_{FS} = \frac{\sum_{i=1}^{n}(r_m - r_i)}{\sum_{i=1}^{n}|r_m - r_i|}$$

Where, $SK_{FS}$ refers to Forhad-Shorna coefficient of "Rank Skewness", $r_m$ is the rank of the mid-range and $r_i$ is the rank of the i$^{th}$ observation of the dataset. This coefficient of skewness ranges from -1 to +1, and is not affected by any extreme value(s) or irregular distance between two consecutive values.

"Rank Skewness" is particularly based on a concept that symmetry is nothing but the existence of equal number of observation at equal distance from the middle point of a dataset i.e. from mean, median or midrange. Here, the reason behind using mid-range is that, it represents the exact midpoint of a dataset, even for a skewed dataset. So, to be symmetric mean, median, mode and mid-range of a dataset should be same. This is why, we have presented this new concept of measure of central value instead of all existing other measures of center values to measure skewness.

In our suggested new form, ranking is done in such a way that the rank of the observations are not affected by extreme value(s) or outlier, it only considers the number of observation at the left side or at the right side from the middle point (midrange) of a dataset and rank is free from the influence of extreme distance or irregular distance from the middle point of a dataset. It will consider only the ranks of the observations not the magnitude of the observations. During determining "Rank Skewness", the following steps are taken into account,

1. Dataset is ordered from smallest to largest value.
2. Mid-range is determined by taking the average of the smallest and the largest value of the dataset, and then the value of the mid-range is inserted into the original dataset. After inserting the midrange in the dataset the whole dataset is again ordered and ranked in Standard competition ranking ("1224" ranking) method and these rank is considered as $r_i$.
3. Rank of mid-range is termed as $r_m$.

4. Sum of all $(r_m - r_i)$ i.e. $\sum_{i=1}^{n}(r_m - r_i)$ is placed at the numerator.

5. Summation of all $|r_m - r_i|$ i.e. $\sum_{i=1}^{n}|r_m - r_i|$ is placed at the denominator. Finally, the quotient is the measure of the Forhad-Shorna coefficient of "Rank Skewness" i.e. $SK_{FS} = \frac{\sum_{i=1}^{n}(r_m - r_i)}{\sum_{i=1}^{n}|r_m - r_i|}$ .

**2.2 Four Point Summary Graph**

Five point summary is represented by the traditional boxplot which is a graphical tool for representing the scatteredness of a dataset that ultimately may be used to check the skewness. Due to some limitation of boxplot, a new approach of four point summary graph has been suggested which is an alternative to the traditional five point summary boxplot. It is termed as "Four Point Summary Graph".

Our proposed method takes into account hundred percent observation of a dataset. As it is based on four points of a dataset and these are minimum, maximum, median and mid-range, so we can term it as four point summary.

In constructing a four point summary graph, the values of a dataset are plotted horizontally and a horizontal line is drawn. After that, minimum, maximum, median and mid-range are plotted on this line. Here, the reason behind using mid-range is that, it is the exact middle point of a dataset. As median represents the 50% observation of a dataset to its left and the rest 50% observation to its right side, so the position of median compared to mid-range is the salient concept.

If the median lies at the left side of mid-range, the dataset may be considered as positively skewed and if it falls at the right side of mid-range, the dataset is negatively skewed, and if median falls on the mid-range point, the dataset may be considered as a symmetric one.

So the four point summary graph will clearly direct the skewness of a dataset, distance between mid-range and median will determine the amount of degree of skewness but it will not exactly determine the amount of skewness.

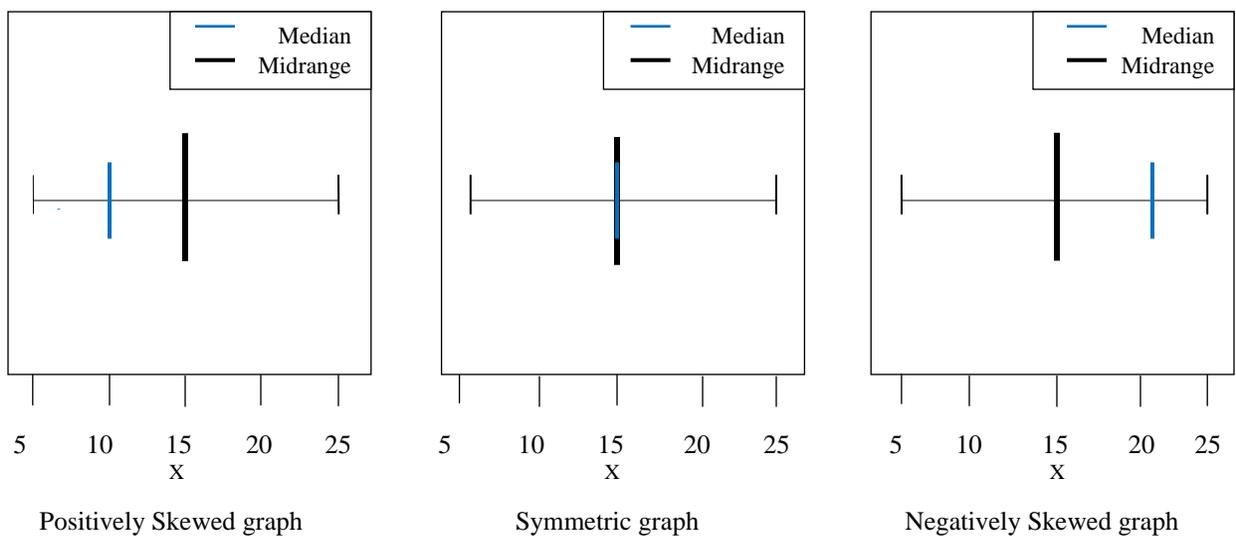

Positively Skewed graph          Symmetric graph          Negatively Skewed graph

Figure 1: Template of "Four Point Summary Graph".

**2.4 Determination of Sample Skewness**

For the statistical analysis regarding this study, firstly, we need dataset of sample skewness. Using Monte Carlo simulation technique with a seed of 2147483647 (one of the largest prime number), we have made data bank each of size 2000000 by generating random number from each of the considered distribution (Normal, positively skewed Gamma (2, 2), positively skewed Weibull (2, 2) and negatively skewed Weibull (10, 4) and, Lognormal distribution). Then, we have resampled from the data bank for each of the above-mentioned distribution for 500000 times using bootstrap method with sample size 10, 20, 30, 40, 50, 60 and 100.

Finally, we have used five formulae for determining skewness for each of these 500000 dataset for all above specified distributions. The used formulae are as follows,

Pearson's second coefficient of skewness (using mean and median) (Pearson), Fisher's moment coefficient of skewness (Moment), Yule's coefficient of skewness (Bowley), Forhad-Adnan measures of skewness using median (FA) and the proposed Rank skewness (FS Rank).

**2.5 Comparison among Different Coefficient of Skewness**

We have considered five different formulae for measuring the coefficient of skewness. "Rank Skewness" has been proposed to resolve some problems in measuring coefficient of skewness as the traditional formulae are affected by that problems. We have compared the efficiency of the proposed Rank skewness with the existing methods as mentioned above.

For comparison, firstly, we have considered standard deviation, mean deviation from mean, mean deviation from median of 500000-sample skewness using each of the above mentioned form for all the considered probability distributions. The smaller the deviation, the better the formula.

Secondly, we have considered the form of the coefficient of population skewness for all the five distribution for comparison. For each of the above-mentioned distributions, we have generated random number of size 2000000 by Monte Carlo Simulation technique and have determined the coefficient of population skewness using respective formulae for the mentioned distributions. Then we have determined the coefficient of skewness by the above-mentioned formulae. The closer the value of the coefficient of skewness using different formulae to the value of the coefficient of population skewness, the better the formulae.

But the coefficient of population skewness is based on only the moment method, this is why; the results obtained by first method may be more preferable for comparison.

**3. Explanation of Comparison**

According to the above mentioned method of comparison, the following results have been observed from our analysis. For Lognormal distribution, based on standard deviation, the proposed "Rank Skewness" performs best (having smallest variation) compared to the other four formulae for all the specified sample sizes. This is true for mean deviation from mean and mean deviation from median also.

For Gamma (2, 2) distribution, the proposed "Rank Skewness" shows smaller variation with the increase of sample size and become best for sample size 100 in terms of standard deviation. For mean deviation from mean and mean deviation from median, similar types of result were observed. For this distribution, we have noticed that these deviation is smallest for sample size 50.

For Normal distribution, the variation is largest in terms of all the three types of measure. In this case, Forhad-Adnan skewness produces the smallest variation.

For negatively skewed Weibull (10, 4) distribution, Forhad-Adnan skewness gives the smallest variation.

The result obtained for positively skewed Weibull (2, 2) distribution is presented in the following tables.

**Table I: Standard deviation of sample skewness for (Weibull (2, 2))**

| Form / Size | Pearson | Moment | Bowley | FA | FS Rank |
|---|---|---|---|---|---|
| 20 | 0.3156221 | 0.3591132 | 0.1685702 | 0.1393499 | 0.2384625 |
| 30 | 0.2818825 | 0.3378248 | 0.1472222 | 0.1225327 | 0.2382569 |
| 40 | 0.2583495 | 0.3168840 | 0.1321438 | 0.1115257 | 0.2313313 |
| 50 | 0.2411682 | 0.2981643 | 0.1215104 | 0.1036155 | 0.2219930 |
| 60 | 0.22788475 | 0.28159314 | 0.11343257 | 0.09761625 | 0.21182796 |
| 100 | 0.19271999 | 0.23411514 | 0.09278911 | 0.08206447 | 0.17333845 |

**Table II: Mean deviation from median of sample skewness for (Weibull (2, 2))**

| Form / Size | Pearson | Moment | Bowley | FA | FS Rank |
|---|---|---|---|---|---|
| 20 | 0.2584922 | 0.2796075 | 0.1376038 | 0.1139764 | 0.2040877 |
| 30 | 0.2307083 | 0.2634700 | 0.1194658 | 0.1002047 | 0.1996434 |
| 40 | 0.21139759 | 0.24749531 | 0.10698664 | 0.09118233 | 0.18962565 |
| 50 | 0.19726332 | 0.23319758 | 0.09832008 | 0.08468119 | 0.17966146 |
| 60 | 0.18627364 | 0.22070039 | 0.09164762 | 0.07972309 | 0.16923472 |
| 100 | 0.15713514 | 0.18380717 | 0.07500294 | 0.06687901 | 0.13466431 |

**Table III: Mean deviation from mean of sample skewness for (Weibull (2, 2))**

| Form / Size | Pearson | Moment | Bowley | FA | FS Rank |
|---|---|---|---|---|---|
| 20 | 0.2597178 | 0.2831658 | 0.1389009 | 0.1146323 | 0.2055509 |
| 30 | 0.2316169 | 0.2659577 | 0.1206995 | 0.1006655 | 0.2009291 |
| 40 | 0.21214447 | 0.24922255 | 0.10812903 | 0.09154932 | 0.19275251 |
| 50 | 0.19780531 | 0.23446461 | 0.09933635 | 0.08494124 | 0.18296572 |
| 60 | 0.18669430 | 0.22167637 | 0.09258698 | 0.07993267 | 0.17240938 |
| 100 | 0.15724099 | 0.18429422 | 0.07570631 | 0.06693861 | 0.13779070 |

Checking the result of all the above tables, it may be noticed that Rank skewness produced better result than Fisher's moment measure of skewness and Pearson's measure of skewness. It may also be noticed that, Bowley's measure of skewness perform better than existing Pearson's method or Fishers Moment method. But as this method is based on only the middle most fifty percent observation and the skewness is determined based on only three observations ($25^{th}$ percentile point observation, $50^{th}$ percentile point observation and $75^{th}$ percentile point observation) so this method is less dependable.

Variation for all the formulae for all the distribution under study decreases with the increase of sample size.

Using the second method of comparison, we have noticed that, for all the distributions so far we have studied, the value of the coefficient of skewness based on moment method provides approximately the same value of population skewness. This may happened as both the population skewness and sample skewness have used the same formulae. Our proposed "Rank skewness" gives sometimes the second closest or the third closest value to the value of population skewness for all the considered distribution.

### 4. Application in Real Life Problem

Here we have considered three real life problems taken from Daniel (2007) and Devore (2000).

The first problem is about the nutritional status of 107 subjects for determining the efficiency of BCG (Bacillus Calmette-Guerin) vaccine in preventing tuberculous meningitis. Their nutritional status scores are given below. We have detected outlier by using EUPP method Hossain (2007) and found three outliers on the right tail of the dataset. All the datasets with their graphical representation and proposed four point summary graph are as follows,

| Dataset-1 | 73.3, 80.5, 50.4, 64.8, 74.0, 72.8, 72.0, 59.7, 90.9, 76.9, 71.4, 45.6, 77.5, 60.6, 67.5, 54.6, 71.0, 66.0, 71.0, 74.0, 72.7, 73.6, 97.5, 89.6, 70.5, 78.1, 84.6, 92.5, 76.9, 76.9, 59.0, 82.4, 56.8, 83.0, 76.5, 72.6, 65.9, 70.0, 130.0, 76.9, 88.2, 63.4, 123.7, 65.6, 80.2, 84.7, 82.6, 76.5, 80.6, 72.3, 99.6, 80.7, 73.3, 77.4, 68.1, 74.6, 70.5, 58.8, 93.7, 61.3, 76.9, 78.2, 85.4, 72.2, 100.0, 55.7, 79.3, 109.0, 84.4, 76.4, 86.4, 67.7, 74.0, 92.3, 76.9, 64.5, 88.7, 72.4, 65.7, 73.6, 79.6, 64.1, 76.9, 68.6, 73.2, 66.3, 70.0, 91.9, 55.5, 100.0, 79.6, 72.7, 78.1, 68.3, 65.9, 74.0, 67.3, 66.3, 96.0, 73.8, 70.0, 50.5, 73.0, 55.0, 80.0, 84.0, 50.9 |

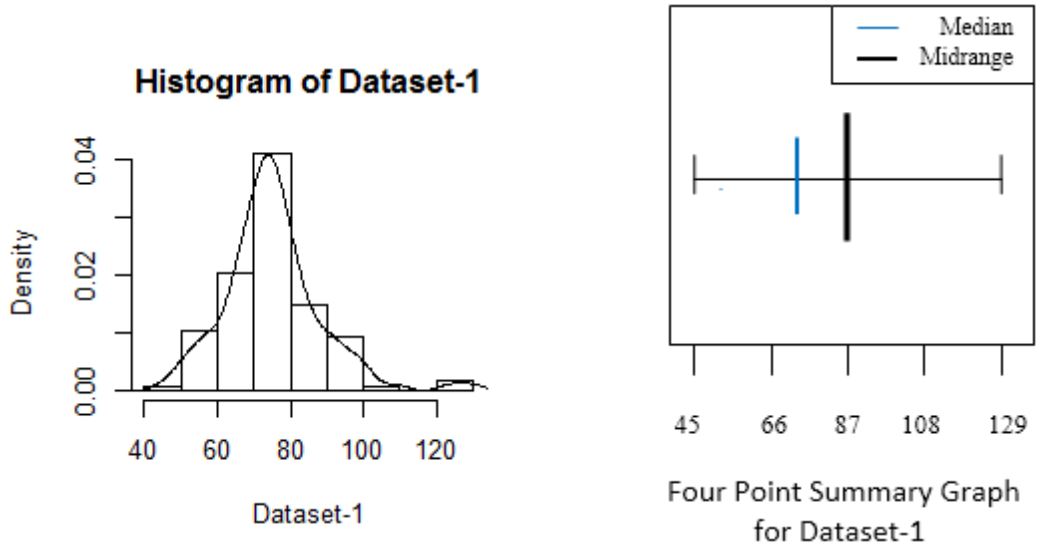

There are outliers in the dataset, which may affect the measures of center value and measures of asymmetry. Here, we have calculated the central value by using mean, median and mid-range, and have obtained the coefficient of skewness using the above-mentioned forms. By comparing with all the above specified forms, the proposed "Rank Skewness" performs well in measuring skewness for this dataset.

**Table V: Value of the coefficient of skewness.**

| Method<br>Dataset | Pearson<br>(-3 to +3) | Moment<br>(-3 to +3) | Bowley<br>(-1 to +1) | FA<br>(-1 to +1) | Rank<br>(-1 to +1) |
|---|---|---|---|---|---|
| Dataset-2 | 0.35118 | 0.993362 | 0.042471 | 0.16678 | 0.93809 |

Following two problems have been taken from Devore (2000).

The problems are on the radon concentration and childhood cancer. The first sample is on the houses where a child diagnosed with cancer had been residing and the second sample is on the houses having no recorded cases of childhood cancer. We have detected outlier by using EUPP method as suggest by Hossain (2007) and found two outliers on the right tail of the first sample and three outliers on the right tail of the second sample.

Following are the datasets with their histogram, frequency curve and four point summary graphs.

| Dataset-2 | 5, 9, 7, 9, 3, 8, 6, 10, 15, 11, 13, 13, 12, 15, 11, 18, 16, 16, 10, 15, 11, 18, 11, 10, 17, 16, 18, 22, 27, 23, 21, 20, 21, 22, 23, 33, 34, 38, 39, 45, 57 |
|---|---|
| Dataset-3 | 9, 5, 7, 6, 8, 3, 9 7, 6, 7, 8, 9, 9, 3, 11, 12, 12, 17, 11, 13, 17, 11, 11, 11, 14, 29, 29, 24, 29, 24, 21, 29, 21, 38, 39, 53, 33, 55, 85 |

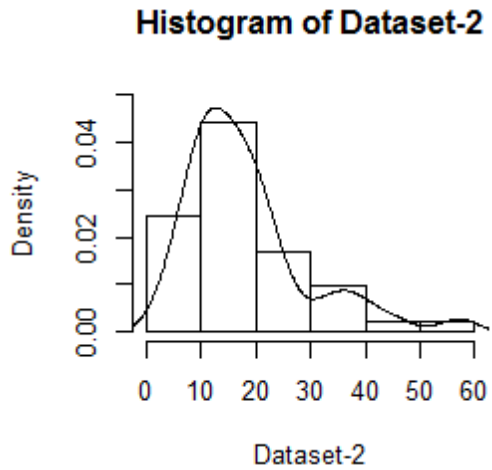
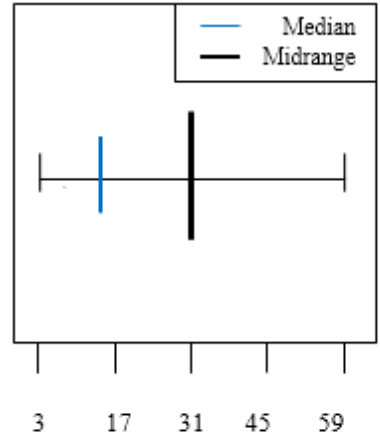

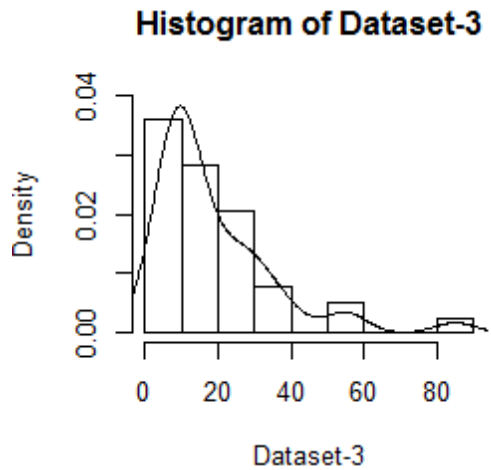
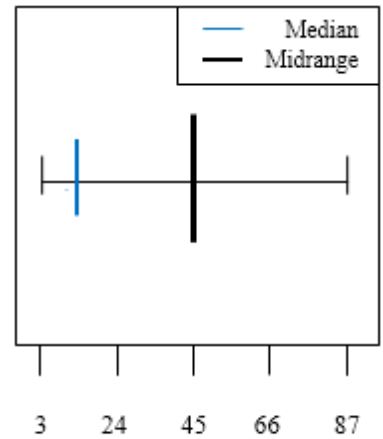

There are outliers in both the datasets, which may affect the measures of central tendency and measures of asymmetry. In both cases, we have obtained the central value by using mean, median and mid-range, and have determined the coefficient of skewness by using the above mentioned forms. From the following table, it is evident that, the proposed "Rank Skewness" performs well in measuring skewness compared to all the above specified forms for these dataset.

Table V: Value of the coefficient of skewness.

| Form<br>Dataset | Pearson<br>(-3 to +3) | Moment<br>(-3 to +3) | Bowley<br>(-1 to +1) | FA<br>(-1 to +1) | Rank<br>(-1 to +1) |
|---|---|---|---|---|---|
| Sample-1 | 0.591003 | 1.428262 | 0.0909091 | 0.283951 | 0.937685 |
| Sample-2 | 1.263187 | 1.917903 | 0.611111 | 0.644342 | 0.985775 |

To calculate the proposed "Rank Skewness", we have suggested midrange as it works well in measuring the coefficient and direction of skewness.

## 5. Conclusions

Skewness is a term, which arises due to the asymmetry of a dataset. To draw a conclusion based on the equality of mean and equality of variance, the prior task is to test the significance and equality of sample skewness. But in measuring the coefficient of skewness, a number of shortcomings are noticed with the traditional forms. Some of the shortcomings may occur due to the presence of a few extreme value(s), largest or smallest distance from the middle point of a dataset and distance between two consecutive numbers. To resolve all these shortcomings we have proposed a new measure of skewness here which is very easy to use, we call it Rank Skewness.

The performance of different measures of skewness have been compared and it has been observed that the proposed "Rank Skewness" performs better than other forms specially for skewed distribution.

Five point summary (traditional boxplot) is a graphical tool that are used to detect the existence and direction of skewness. We have suggested an alternative graph over the existing five point summary (traditional boxplot). We termed it as four point summary graph. This graph is based on all observations and is simpler. It is also not affected by extreme values and gives better results than the traditional one. So, proposed four point summary graph may be considered as an alternate to the five-point summary.